\documentclass[10pt,conference]{IEEEtran}
\IEEEoverridecommandlockouts
\usepackage{cite}
\usepackage{amsmath,amssymb,amsfonts}
\usepackage{algorithmic}
\usepackage{graphicx}
\usepackage{textcomp}
\usepackage{xcolor}
\usepackage{algorithmic}
\usepackage{ifthen}
\usepackage{url}
\usepackage{booktabs}
\usepackage[export]{adjustbox}

\def\BibTeX{{\rm B\kern-.05em{\sc i\kern-.025em b}\kern-.08em
    T\kern-.1667em\lower.7ex\hbox{E}\kern-.125emX}}

\begin{document}
\title{Do Communities in Developer Interaction Networks align with Subsystem Developer Teams? An Empirical Study of Open Source Systems}

\author{
\fontsize{8.3}{9.5}\selectfont
\centering
\IEEEauthorblockN{Usman Ashraf}
\IEEEauthorblockN{Christoph Mayr-Dorn}
\IEEEauthorblockN{Atif Mashkoor}
\IEEEauthorblockN{Alexander Egyed}
\IEEEauthorblockA{Johannes Kepler University, Austria\\
firstname.lastname@jku.at}
\and
\IEEEauthorblockN{Sebastiano Panichella}
\IEEEauthorblockA{Zurich University of Applied Science, Switzerland\\
sebastiano.panichella@zhaw.ch}\\
[0.9cm]  
}

\maketitle

\begin{abstract}
Studies over the past decade demonstrated that developers contributing to open source software systems tend to self-organize in ``emerging'' communities. This latent community structure has a significant impact on software quality. While several approaches address the analysis of developer interaction networks, the question of whether these emerging communities align with the developer teams working on various subsystems remains unanswered. Work on socio-technical congruence implies that people that work on the same task or artifact need to coordinate and thus communicate, potentially forming stronger interaction ties. Our empirical study of 10 open source projects revealed that developer communities change considerably across a project's lifetime (hence implying that relevant relations between developers change) and that their alignment with subsystem developer teams is mostly low. However, subsystems teams tend to remain more stable. These insights are useful for practitioners and researchers to better understand developer interaction structure of open source systems.

\end{abstract}


\section{Introduction} \label{sec:intro}

Global software development is carried out by developers located in various parts of the globe. They face a number of challenges in relation to communication and coordination due to the distances involved in three dimensions -- geographical, temporal, and socio-cultural \cite{DBLP:journals/cacm/ConchuirAOF09}.  Open source software development is considered as a successful example of large scale global software development \cite{DBLP:conf/euromicro/GaughanFS09}. In open source development environment, software systems do not have a pre-assigned organizational structure as developers can contribute to any part of the system. However, developer communities are self-organized within development teams~\cite{Bird:fse08}. Studies have shown that the developer communities organization has a significant impact on software quality \cite{DBLP:conf/icse/NagappanMB08,DBLP:conf/sigsoft/MeneelyWSO08,DBLP:conf/sigsoft/PinzgerNM08,DBLP:journals/tse/CataldoH13}.

Motivated by the aforementioned insights, researchers have built automated tools analyzing developers interaction data from various sources, such as version control systems, mailing lists, or issue trackers \cite{DBLP:journals/ijitwe/Lopez-FernandezRGH06, Bird:fse08,Jermakovics:2011:MVD:1984642.1984647,DBLP:conf/icse/MeneelyW11,PanichellaBPCA14,7194606}, to investigate how \textit{``emerging development teams''} are formed in open source projects. However, a few researchers observed how the emerging \textit{``community structure''} aligns with \textit{``subsystem developer teams''}. Bird et al. \cite{Bird:fse08}, for example, hypothesized that files edited by developers of the same community are placed ``closer together'' than the files edited by randomly picked developers. Results, however, were inconclusive.

We argue that observing the alignment of developer communities with the system architecture can be more accurately studied at the level of coarse-granular subsystems, i.e., the components of a system, rather than analyzing these aspects at the level of fine-grained artifacts. In this regard, the work of Lenarduzzi et al. \cite{10.1145/3234152.3234191}, for example, is important as they point out the potential negative impact of team independence at the subsystem (microservice) level.
Additionally, Nagappan et al. \cite{DBLP:conf/icse/NagappanMB08} have shown that organizational metrics are the top bug proneness predictor in an industrial setting. Given the lack of explicit organizational structures in open source systems, we suggest to utilize implicit structures (i.e., communities)\footnote{We define a community as the set of developers that share responsibility or interest such as working on the same subsystem(s)~\cite{7194606}.}.

\begin{figure}[ht]
    \centering
    \includegraphics[width=01.0\columnwidth]{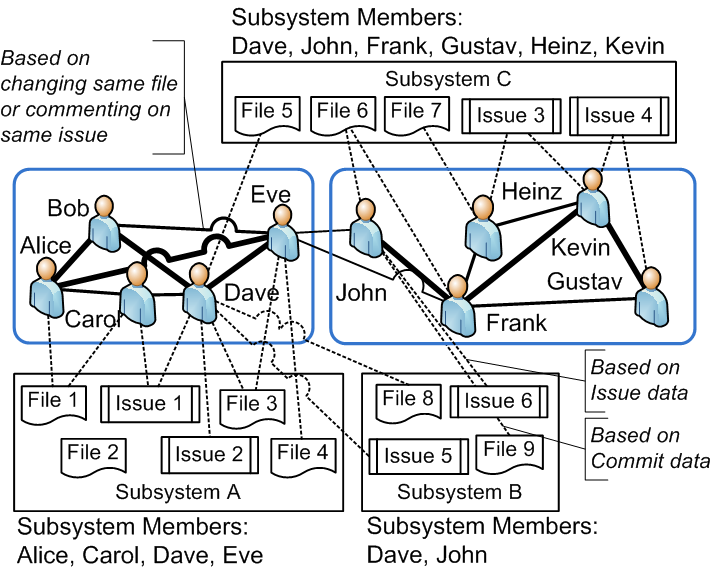}
    \caption{Example developer community to subsystem alignment.}
    \label{fig:examplecommunityalignment}
\end{figure}

Take the excerpt from a fictive project in Fig. \ref{fig:examplecommunityalignment} as an example.
The \emph{developer interaction network} on the top exhibits edges based on  
being involved in the same issue via activities such as reviewing or commenting. A community detection algorithm assigns developers to communities (blue borders) based on the developers' interaction intensity (indicated by line thickness). Developers who interact often tend to form communities.
The lower part of Fig. \ref{fig:examplecommunityalignment} depicts files and issues belonging to three subsystems A, B, and C. Subsystem team members are then those developers that changed an artifact belonging to that subsystem, respectively, are active in an issue referring to this subsystem. Note that subsystem member sets may overlap as in open source projects key developers are often involved in multiple subsystems. In our example, developer Dave is a member of all three subsystems, Carol is a member of subsystem A only.
In this example, when comparing the members of subsystem B (i.e., Dave and John), we notice that they are members of different communities and there is no edge between them. This could be a warning signal about poor communication and coordination between the two.

To shed light into how communities and subsystems are aligned, we empirically investigate in this paper, in the context of 10 open source projects, how communities emerge and change over time -- e.g., how developers join and leave sub-communities -- and the extent to which these community patterns match the subsystems evolution.
We find that developer communities change considerably across a project's lifetime (hence implying that relevant relations between developers change) while subsystem developer teams (SDTs) remain comparatively stable. Overall, the community alignment with SDTs is often low, which implies that developers maintain significant communication ties with developers outside their (subsystem) work scope. We hypothesise that such an interaction network independent from subsystems emerges from the need to remain robust against the disruption of leaving developers and quick onboarding of new members.  

The primary contribution of this paper is an empirical study investigating the evolution of developer communities and their alignment with the subsystems developer teams (Section \ref{sec:dataanalysisprocedure}). The secondary contributions of this paper are: i) a technique for measuring alignment (i.e., overlap) between subsystems and developer communities (Section \ref{sec:dataanalysisprocedure}(C)), and ii) a technique for determining developer communities evolution  (Section \ref{sec:dataanalysisprocedure}(D)).

The paper is structured as follows. Section \ref{sec:studydesign} introduces the research questions. Section \ref{sec:datagathering} describes the applied data gathering method and the resulting evaluation data set. 
Section \ref{sec:dataanalysisprocedure} explains our community-SDT alignment and evolution measurement technique 
 with Section \ref{sec:results} presenting the results. We discuss the obtained results and their implications in Section \ref{sec:discussion}. 
Section \ref{sec:sota} compares this study to state of the art approaches before Section \ref{sec:conclusions}, which concludes the paper with a summary and an outlook on the future work.

\section{Study Design} 
\label{sec:studydesign}

Developers that work on  the  same task, artifact, or subsystem need to coordinate and thus communicate in order to avoid incomplete change propagation, rework, or duplicate work. We therefore would expect them to form stronger interaction ties \cite{DBLP:conf/oss/SyeedH13} than developers working in different subsystems. We hypothesize that these developer interactions give rise to a community structure.
Therefore, the \emph{goal} of this study is to obtain first insights into the alignment of sub communities within open source projects with the systems' structure.

 The \emph{motivation} behind this study is to determine the extent to which emerging open source developer communities reflect the subsystem structure, proposing mechanisms for studying such communities aspects.
 The \emph{perspective} is of practitioners and researchers that could leverage such mechanisms to identify subsystems where developers coordinate insufficiently.

The \emph{context} of this study comprises commits, issues, source code (folder) structure, and the conceptual and structural links among the data. The selected 10 open source projects are part of our publicly available and published dataset \cite{DBLP:conf/msr/AshrafMEP20}. 

\begin{figure*}[htb]
    \includegraphics[width=0.95\textwidth, height= 7cm]{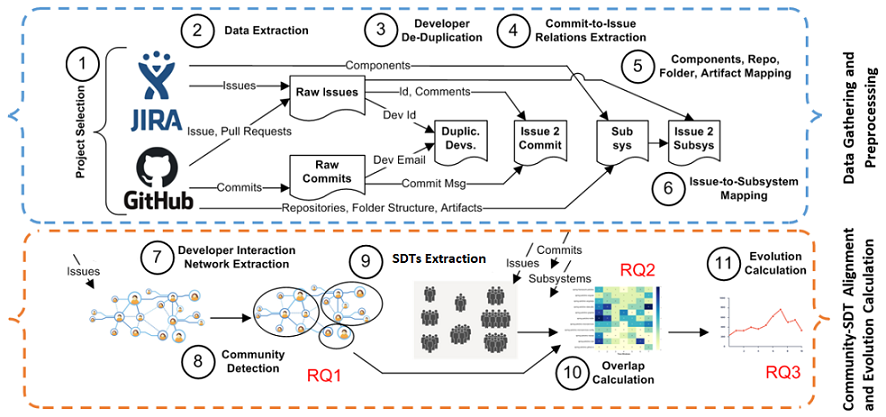}
    \caption{Steps of the study approach.}
    \label{fig:study}
\end{figure*}

\textbf{Research Questions}: \\
In this work, we investigate the following research questions:\\ 
\textbf{RQ1}: \textbf{\emph{To what extent can we identify well defined developer communities across the investigated projects' lifetime?}}
Our goal is to propose a systematic approach able to detect emerging communities by analyzing the \emph{developer interaction network} generated from issue involvement and commits  linked  to  those  issues (see Fig.~\ref{fig:examplecommunityalignment}). 
    We are then interested in knowing how often and how many well defined communities we detect.\\
\textbf{RQ2}: \textbf{\emph{Do the developers active in the same subsystem emerge in the same development communities?}}  
    In other words, measuring the overlap of the SDTs with the detected communities allows us to identify subsystems where the communication among developers occurs mostly within the subsystem. 
\\
\textbf{RQ3}:\textbf{\emph{ How stable are the detected communities across time compared to the SDTs?}}  While RQ1 and RQ2 investigate \emph{developer interaction network} and its alignment with SDTs at separate snapshots in time, here, we inspect the stability of the detected communities and the SDTs across time. We analyse whether a pair of developers that belong to the same community in one time window also belong to it in the subsequent time window. Likewise, we are interested to know whether a pair of SDTs that emerge in the same developer community does so in the subsequent time window.

Overall, answering these questions provides insights into whether subsystems in open source projects represent decoupled work scopes that result only in limited coordination overhead compared to work coordination within a subsystem (as measured by interaction ties).

\section{Data Gathering and Preprocessing} \label{sec:datagathering}

\subsection{Open Source Project Selection}

The project selection (Fig.~\ref{fig:study} (1)) is the first step of our study approach. We select the candidate project according to the following requirements:

\textbf{Subsystem Structure}
We manually selected projects which exhibit a non-trivial (i.e., at least 10 subsystems) and clear subsystem structure to avoid introducing potential bias by splitting the system into subsystems that do not reflect the real underlying decomposition. 

We manually inspected the projects on  Apache's Jira server\footnote{\url{https://issues.apache.org/jira/}} that exhibited a significant number of issues with a \emph{component} property set and where the top level source code folders (hosted on Github) closely (or identically) matched those component names. We interpreted those components/folders as subsystems.

We avoid selecting projects from a single mechanism for structuring subsystems by also selecting Github projects that manage code across multiple repositories (multi-repo), i.e., one repository per subsystem, as our selected Jira projects are mono-repo, i.e., one folder per subsystem.

\textbf{Developers}
Further selection criteria included a minimum of 40 participating developers over the project's life time to guarantee that even with heavy fluctuation of developers, that there are sufficiently many developers to form an interaction network with sub-communities. Note that for our selected projects the number of developers are well over 40.

\textbf{Commits and Issues}
Furthermore, we filtered out projects with less than 1500 commits or 200 issues. This ensures that a relation between two developers based on being active in the same issue or  committing an  artifact  update  linked  to  the  same  issue may occur sufficiently often to signify a meaningful relation between the developers and not just an one-off occurrence.

This study focuses on 10 projects listed in Table~\ref{tab:projects}. The projects are limited in number and size as (i) there is a manual processing effort required in step 1 (Fig.~\ref{fig:study})  
and (ii) non-negligible manual effort is necessary to investigate folder structure and confirm developer deduplication.

\subsection{Data Extraction and Storing}

Having identified 10 projects under investigation, we extracted commit and issue information (Fig.~\ref{fig:study} (2)).
We used \emph{Perceval} \cite{Duenas:2018:PSP:3183440.3183475} to extract commit and issue information from Github  
and the Jira python client (\emph{Perceval} lacks to provide our desired Jira issue information) for extracting data from Apache Jira.  
Both these tools provide data in the JSON format.

Commits, the involved artifacts, and issue details make up the core information of our data set.
In our datamodel, an issue may represent a Jira issue, a Github issue, or a Github pull request, or even a combination thereof whenever, for example, a pull request references a Jira issue.

A developer participates in a project in various ways: for example, from committing code changes, to commenting on issues, to reviewing pull requests.
We introduce two \emph{Involvement} types to harmonize activities across Jira and GitHub. \emph{Contributing} is equal to committing artifacts (i.e., as indicated by a commit linked to an issue); \emph{Informative} describes input to an issue such as having reported it, commented on it, or having reviewed artifacts. We encode these two \emph{types} of actions as integers of value 3 and 2, respectively, to reflect the amount of effort behind the activities. Before continuing with these values we performed a value sensitivity analysis  by assigning them values (4,2) and (1,1) respectively. We observed little change in the detected community structure with relatively less quality. More details of community structure and its quality metric are in the Section \ref{sec:dataanalysisprocedure}(A).

\begin{table*}[!ht]
\caption{Overview of the ten analysed projects.}
\centering
\resizebox{\textwidth}{!}{%
\begin{tabular}{llllrrrrrcccc}\hline
\multicolumn{1}{|l|}{\textbf{Project}} & \multicolumn{1}{c|}{\textbf{Type}} & \multicolumn{1}{l|}{\textbf{\begin{tabular}[c]{@{}l@{}}Programming\\ Language(s)\end{tabular}}} & \multicolumn{1}{c|}{\textbf{\begin{tabular}[c]{@{}c@{}}Time Period \\ {[}Months{]} (From-To)\end{tabular}}} & \multicolumn{1}{r|}{\textbf{Commits}} & \multicolumn{1}{r|}{\textbf{Devs}} & \multicolumn{1}{r|}{\textbf{Subsys}} & \multicolumn{1}{r|}{\textbf{Arts}} & \multicolumn{1}{r|}{\textbf{Issues}} & \multicolumn{1}{c|}{\textbf{\begin{tabular}[c]{@{}c@{}}Arts Linked to \\ Subsys {[}\%{]}\end{tabular}}} & \multicolumn{1}{c|}{\textbf{\begin{tabular}[c]{@{}c@{}}Issues Linked To \\ Subsys {[}\%{]}\end{tabular}}} & \multicolumn{1}{c|}{\textbf{\begin{tabular}[c]{@{}c@{}}Issues Linked To \\ Commits {[}\%{]}\end{tabular}}} & \multicolumn{1}{c|}{\textbf{\begin{tabular}[c]{@{}c@{}}Commits Linked To \\ Issues {[}\%{]}\end{tabular}}} \\ \hline
lagom                                  & Multi-Repo                         & Scala, Java                                                                                     & 44 (Mar 16-Nov 19)                                                                                           & 6089                                  & 540                               & 17                                   & 27490                              & 3381                                 & 100                                                                                                     & 100                                                                                                       & 59                                                                                                         & 62                                                                                                         \\
nameko                                 & Multi-Repo                         & Python                                                                                          & 85 (Sep 12-Nov 19)                                                                                           & 3861                                  & 233                               & 14                                   & 960                                & 867                                  & 100                                                                                                     & 100                                                                                                       & 65                                                                                                         & 93                                                                                                         \\
kumuluz                                & Multi-Repo                         & Java, JavaScript                                                                                & 54 (May 15-Nov 19)                                                                                           & 2487                                  & 82                                & 19                                   & 2083                               & 274                                  & 100                                                                                                     & 100                                                                                                       & 45                                                                                                         & 22                                                                                                         \\
jhipster                               & Multi-Repo                         & Java, JavaScript                                                                                & 71 (Nov 13-Nov 19)                                                                                           & 10391                                 & 767                               & 11                                   & 7146                               & 3267                                 & 100                                                                                                     & 100                                                                                                       & 64                                                                                                         & 59                                                                                                         \\
networknt                              & Multi-Repo                         & Java, JavaScript                                                                                & 38 (Sep 16-Nov 19)                                                                                           & 5369                                  & 110                               & 12                                   & 16127                              & 2005                                 & 100                                                                                                     & 100                                                                                                       & 74                                                                                                         & 36                                                                                                         \\
flume                                  & Mono-Repo (Jira)                   & Java                                                                                            & 115 (Jun 10- Jan 20)                                                                                         & 2623                                  & 1199                              & 16                                   & 3403                               & 3638                                 & 51                                                                                                      & 57                                                                                                        & 29                                                                                                         & 86                                                                                                         \\
stanbol                                & Mono-Repo (Jira)                   & Java                                                                                            & 107 (Nov 10-Nov 19)                                                                                          & 6953                                  & 149                               & 21                                   & 17469                              & 1490                                 & 54                                                                                                      & 80                                                                                                        & 47                                                                                                         & 62                                                                                                         \\
falcon                                 & Mono-Repo (Jira)                   & Java                                                                                            & 87 (Nov 11- Mar 19)                                                                                          & 2556                                  & 190                               & 19                                   & 7362                               & 2760                                 & 56                                                                                                      & 38                                                                                                        & 47                                                                                                         & 63                                                                                                         \\
tika                                   & Mono-Repo (Jira)                   & Java                                                                                            & 154 (Mar 07-Jan 20)                                                                                          & 5848                                  & 1327                              & 21                                   & 6509                               & 3335                                 & 71                                                                                                      & 65                                                                                                        & 46                                                                                                         & 67                                                                                                         \\
openjpa                                & Mono-Repo (Jira)                   & Java                                                                                            & 164 (May 06- Jan 20)                                                                                        & 7215                                  & 831                               & 27                                   & 10696                              & 2849                                 & 54                                                                                                      & 73                                                                                                        & 57                                                                                                         & 68                                                                                                           \\
[1pt] \hline
\end{tabular}%
}
    \label{tab:projects}
\end{table*}

\subsection{Developer De-Duplication}

When recording a developer's involvement across commits and issues, we need to take care of situations where a developer uses multiple accounts (email-ids) on Github. More importantly, a developer's id on Jira does not match the Github account id of the same developer as Jira does not use email addresses as a part of a user id while Github does. This will result in inconsistencies as one developer will appear as multiple individuals in the data set, hence the need for deduplication (Fig.~\ref{fig:study} (3)). We used the \emph{Dedupe} library~\cite{dedupe}, which uses machine learning to perform deduplication. It has an accuracy up to 95\% \cite{dedupedoc}. 
More details of the training process are available in our dataset \cite{DBLP:conf/msr/AshrafMEP20}.

\subsection{Extraction of Commit-To-Issue Relations}

Github commit messages often refer to issues to identify the purpose of the changes. We manually inspected each project to identify what patterns developers tend to apply for referring to issue-id and applied that pattern when parsing the commit messages (Fig.~\ref{fig:study} (4)).
The following are two sample commit messages from our data referring to issue-id: FLUME-3311 and kumuluz\#115 respectively.
\begin{itemize}
  \item \emph{``FLUME-3311 Update User Guide In HDFS Sink''}
  \item \emph{``Merge pull request \#115 from Jamsek-m/master''}
\end{itemize}

We applied the same mechanism to pull requests. Whenever a pull request identifies a Github issue, we merge the pull request and the Github issue into one dataset issue and link all commits that are part of the pull request to it. This is particularly helpful when commit messages lack a reference to an issue.

For the Jira-backed projects, we regularly find pull requests and commits without a reference to a Github issue but instead a reference to a Jira issue.
 Whenever a pull request identifies a Jira issue, we subsequently merged them into a single issue in our dataset (we never encountered multiple pull requests referencing the same issue) and link all involvements from the pull request and Jira issue to that merged Issue. We proceed likewise whenever we find a Github issue referencing (in the comments or title) a Jira issue.

\subsection{Mapping Subsystems}

For those projects that utilize Apache Jira for issue management, aside from mapping Jira to Github users, we also need to map a Jira project's components to the corresponding folders on Github (Fig.~\ref{fig:study} (5)).
We manually mapped each component of the five Jira projects to its corresponding Github folder or multiple folders where necessary. We link all those folders, which do not map to a component to the main system  (we store main system as a subsystem in our data model).

As briefly mentioned above, the mapping for Github multi-repo projects is straight forward: each repository in the project becomes a subsystem. 
Furthermore, we store the list of folders that make up a subsystem. This allows us to relate all artifacts to a unique subsystem.

\subsection{Linking Issues to Subsystems}
Linking issues to subsystems (Fig.~\ref{fig:study} (6)) works similar to relating artifacts to subsystems. Jira issues exhibit a \emph{Component} property that identifies all affected subsystems (potentially multiple). Github multi-repo projects provide separate issue lists for each repository, thus we link those to subsystems unambiguously.

\subsection{Dataset Overview}\label{sec:datasetoverview}

Table \ref{tab:projects} provides an overview of the 10 chosen open source projects. The columns \emph{Type} and \emph{Programming Language(s)} report the repository structure and languages of the project, respectively.  
The column \emph{Time Period} describes the overall time in months in which we extracted commits and issues. All subsequent rows report values from this time period. 
The \emph{Commits} column reports how many commits are made in total in a project.  
Recall that in multi-repo projects, commits per subsystem belong to one repository, thus the reported numbers are the sum of commits over all repositories in such a project. In case of mono-repo projects, all commits come from a single, main repository.
The \emph{Devs} column provides the total number of developers in each project. We consider any person a developer who is \emph{Contributing} or \emph{Informative} in at least one subsystems.
\emph{Subsys}, \emph{Arts}, and \emph{Issues} columns report the number of subsystems, total source code artifacts, and issues in a project, respectively.

The \emph{Arts linked to Subsys} column shows the percentage of artifacts belonging to subsystems. 
In a Mono-Repo system each folder has its link to a subsystem or to the main system. 
So in this case, the column shows the percentage of artifact which are linked to subsystems and not to the main system.
Each repository (subsystem) of a Multi-Repo system has its own Issues, 
thus, the \emph{Issues Linked To Subsys} column shows their 100\% link to subsystem. In a Mono-Repo system, we also include issues from Jira, where they have specified mapping to a subsystem.
As mentioned in the data preprocessing section, commit(s) are made against an issue and issue(s) are linked against commits. The Columns \emph{Issues Linked To Commits} and \emph{Commits Linked To Issues} specify the percentage of such linkage, respectively. Note that as the linkage is not always present e.g., commit  messages  not always  refer  to  issues  to  identify the  purpose  of  the  changes. Therefore, the linkage percentage ranges between 22 to 93.

\section{A Technique for Measuring Community-SDT Alignment and Evolution} \label{sec:dataanalysisprocedure}

\subsection{Developer Interaction Network Extraction}

RQ1 requires building a developer interaction network. We built such a network from developers' involvement in issues such as contributing code or commenting. 
We sliced the project lifetime into time windows of 4 months, as done in close-related  works \cite{DBLP:conf/icse/MeneelyW11,7194606}. We presume that such a time window is sufficiently long to allow developers to repeatedly interact to produce a reliable interaction network while being short enough to capture changes of the interaction network as developers join and leave. 

\textbf{Developer Interaction Network Model}
We model the developer interaction network (Fig.~\ref{fig:study} (7)) as an undirected graph with developers as nodes ($d \in V$) and edges ($e(d_i,d_k) \in E$) between those developers that interact. We add an edge between two developers when the two developers are involved in the same issue, for example, one developer adding a comment to an issue, the other developer committing an artifact update linked to the same issue. The edge's \emph{issue involvement intensity} measures in how many issues the two developers are active in, weighted by their involvement type.

\textbf{Developer Involvement}
Recall that the developer involvement types -- \emph{Contributing} and \emph{Informative} -- carry a weight of 3 and 2, respectively. We consider only the highest weighted involvement per developer and issue $inv_{max}(d,i)$. The intensity for a single issue $i_h$ and pair of developers $d_i$ and $d_j$ (i.e., the per-issue score) is then the lowest of the two scores $min(inv_{max}(d_i,i_h),inv_{max}(d_j,i_h))$. For example, the per-issue score between a contributing developer and an informative one is 2. The overall intensity for a pair of two developers across all their common issues (i.e., the \emph{issue involvement intensity} edge property) is then the sum of per-issue scores ($\sum_k min(inv_{max}(d_i,i_h),inv_{max}(d_j,i_h))$). The \emph{issue involvement intensity} serves as an edge weight.

\textbf{Community Detection} 
The subsequent step after forming the developer interaction network is to discover communities in it. The communities are the natural divisions of network nodes into densely connected subgroups\cite{Newman_2004}.
The developer interaction graph serves as the input to the step of community detection (Fig.~\ref{fig:study} (8)).  
We first remove all developers who have no edges to any other developer. We consider those as not part of any community and retaining them in the graph would negatively influence the community detection results.

For the purpose of community detection, we use
the order statistics local
optimization method (OSLOM)\cite{10.1371/journal.pone.0018961}. This method is also used in close-related work \cite{7194606}. \emph{OSLOM} finds statistically significant communities and has distinct features such as to handle weighted graphs, to form overlapping communities, and to distinguish communities from pseudo-communities. 

\textbf{Community Verification and Validation} The \emph{OSLOM} algorithm identifies the best distribution of graph nodes as communities after calculating the statistical significance (the probability of finding the cluster in a random
null model, i.e. in a class of graphs without community structure) of each community.  
It uses the significance as
a fitness measure to evaluate and include a community candidate in the resulting set of communities. We additionally checked the resulting quality of the community structure using conductance as follows.

\textit{Quantitative Community Quality Metric:}
We apply conductance \cite{DBLP:conf/www/LeskovecLM10}, which measures the probability of having an edge leaving the community. A community with all edges connecting only member nodes has conductance 0 (an isolated community) while a community with no edges amongst its members has conductance of 1 (arguably not really a community at all).

\textit{Qualitative Detected Communities Inspection:} In order to validate that the detected communities accurately reflect real-world developer collaboration, two of the co-authors manually inspected \emph{OSLOM's} detected communities of one of our projects, i.e., \emph{kumuluz}. The project has 82 developers and 54 months of observed developers interactions. We inspected the detected communities for 5 time windows (20 months). We verified each detected community of every window by matching the strength of its collaborative relations compared to the other undetected communities. We also observed the correctness of developers division into communities based on edge weight. 

The resulting communities consists of a subset of developers ($c_k \subset V$) where developers potentially are  members of more than one community (overlapping communities). This is an important property as in developer interaction networks, the most active developers are often connected to two or more communities \cite{DBLP:conf/msr/CanforaCCP11}. In the absence of overlapping communities, a developer will be placed into a single community which will skew the
 subsystem developer overlap calculations.

\textbf{Observed Time Windows}
We apply the community detection algorithm for each time window separately.
We skip time windows with less than ten developers as we do not expect to find meaningful communities among such a small number of developers. Manual inspection of sample windows with less than 10 developers confirmed these expectations. For all these windows the best community structures found by \emph{OSLOM} exhibited high conductance values ($>0.8$). 

\subsection{SDT Extraction} \label{sec:SDT}
Answering RQ2 requires identifying for each subsystem ($s_y \in S$) who are the active developers: the SDT ($SDT(s) \subset V$) (Fig.~\ref{fig:study} (9)). To this end, we simply consider a developer as active if one has at least two \emph{Contributing} or five \emph{Informative} involvements in that particular subsystem within the defined time window. 
This ensures we select only developers that have a longer-running interest in a subsystem and not just include anyone with minimal involvement. We continued with these values i.e., two and five after observing the trend of the number of \emph{Contributing} and \emph{Informative} involvements in the chosen projects.

In contrast to the interaction network community construction, a developer becomes member of an SDT purely based on one's contributions to source code and/or involvement in issues regardless of interactions with other developers. For example, two developers changing the same artifact within the same time window that otherwise are not involved in any issue will end up in the same SDT.

Similar to developer overlapping communities, multiple SDTs may contain the same developer. Especially key developers are often involved in multiple subsystems. At the same time, subsystems may be too small to give raise to a dedicated SDT that focuses only on that single subsystem.
Ultimately, for each developer we define the community membership $cm(d) \in [1,k]$ and SDT membership $SDTM(d) ~ \forall ~ S$.

\subsection{Overlap Calculation}
The second part for answering RQ2 is a metric for measuring how well a SDT matches a developer community (Fig.~\ref{fig:study} (10)). In a project where the community structure represents subsystems, we would expect that members of the same SDT are also members of the same community (low membership heterogeneity). On the other side of the spectrum, we would find a SDT where every member belongs to a different community (high membership heterogeneity).
We measure each SDT's membership heterogeneity using the normalized Shannon entropy \cite{shannon1948mathematical} $mh(SDT(s))=-\sum_k(p_k Ln p_k) / Ln k$ where $p_k$ is the number of developers in $SDT(s)$ being member of community $c_k$ and $Ln$ is the natural logarithm. The normalized Shannon entropy provides a result in the interval $[0,1]$: 1 when $p_k$ is the identical for every $k$ and 0 when $p_k=0$ for all but one $k$. In other words, $mh(SDT(s))=1$ when all SDT members are exactly equally distributed across all communities, and $mh(SDT(s))=0$ when all SDT members are member of the same community.
In our fictive example in Fig.~\ref{fig:examplecommunityalignment} $mhSDT$ for Subsystem A is 0 and $mhSDT$ for Subsystem B is 1. Ideally, we observe minimal/low membership heterogeneity for all subsystems.

Having introduced the general concept of membership heterogeneity, we need to outline an adjustment to $p_k$. Without any adjustment each developer has equal impact. However, this does not properly reflect the different developer types in a typical open source project. There is a small number of key developers that get involved in issues and source code changes across the subsystems, subsequently being well connected and informed. Giving them equal weight as a developer who is focused only on one single subsystem would skew the measure.  
We thus redefine $p_k$ as the weighted sum of developers in $SDT(s)$ being member of community $c_k$ where the weight describes the developer's focus $f(d) \in [0,1]$ across the subsystems (no other adjustment to $mh(SDT(s)$ needs to be made).
The focus on a single subsystem is maximal when the developer is active in a single subsystem, and minimal when involved in every subsystem. To this end, developer focus is calculated as 1 minus the normalized Shannon entropy of the developer's involvement in each subsystem (see Subsection \ref{sec:SDT}); hence $f(d) = 1 + \sum_y(p_y Ln p_y / Ln y$) with $p_y = \sum_{i \in y} inv_{max}(d,i)$.

\subsection{Membership Evolution}

Regarding RQ3, we expect community structure as measured by conductance, and subsystem alignment as measured by SDT membership heterogeneity to change over time. Focusing on community structure, we want to be able to interpret community evolution not only through changes in conductance but also by looking at the external stability of developers (joining and leaving developers) and internal stability (shifting of developers from one community to another in two consecutive time windows $w_t$ and $w_{t+1}$) (Fig.~\ref{fig:study} (11)).
Measuring external stability is simply a matter of tracking which developers appear as nodes in the interaction network graph in one time window but not in the next and vice versa. As the detected communities have no meaningful, intrinsic identity (we just assign them a number as identifier) internal stability requires tracking which developers are in the same community in one time window and remain in the same community (irrespective of the community identifier) in the next window. To this end, we counted how many pairs of developers were in the same community in $w_t$ and are again in the same community in $w_{t+1}$ ($\#sameC$) or were in different communities in $w_t$ and are in different communities again in $w_{t+1}$ ($\#diffC$). The internal stability metric is then $iStab = (\#sameC+\#diffC) / totalPairs$ and yields 1 for full stable communities and 0 when all communities become completely reshuffled. 

Similar to developer community membership evolution, we are interested to know whether the SDTs that overlap with the same community in one time window will overlap with the same community in the subsequent time window (Fig.~\ref{fig:study} (11)). We apply the same calculation as for internal stability above but instead of comparing a pair of developers being in the same or different community again, we measure in $sdtStab$ whether pairs of SDTs tend to overlap with the same community in the subsequent time window again or whether they overlap with different communities.

\section{Results} \label{sec:results}
In this section, we present the detailed results for a single example project \emph{flume} only due to page limitation. We thus provide aggregated numbers across all ten projects and refer for per-project details to our supporting online material (SOM) \cite{SOM}. The SOM includes a database dump of the underlying dataset, the source code used for extracting the raw data, preprocessing, and metric calculation, as well as detailed figures for each project.

\subsection{Answering RQ1}
\begin{figure}[!h]
    \includegraphics[width=1.0\columnwidth]{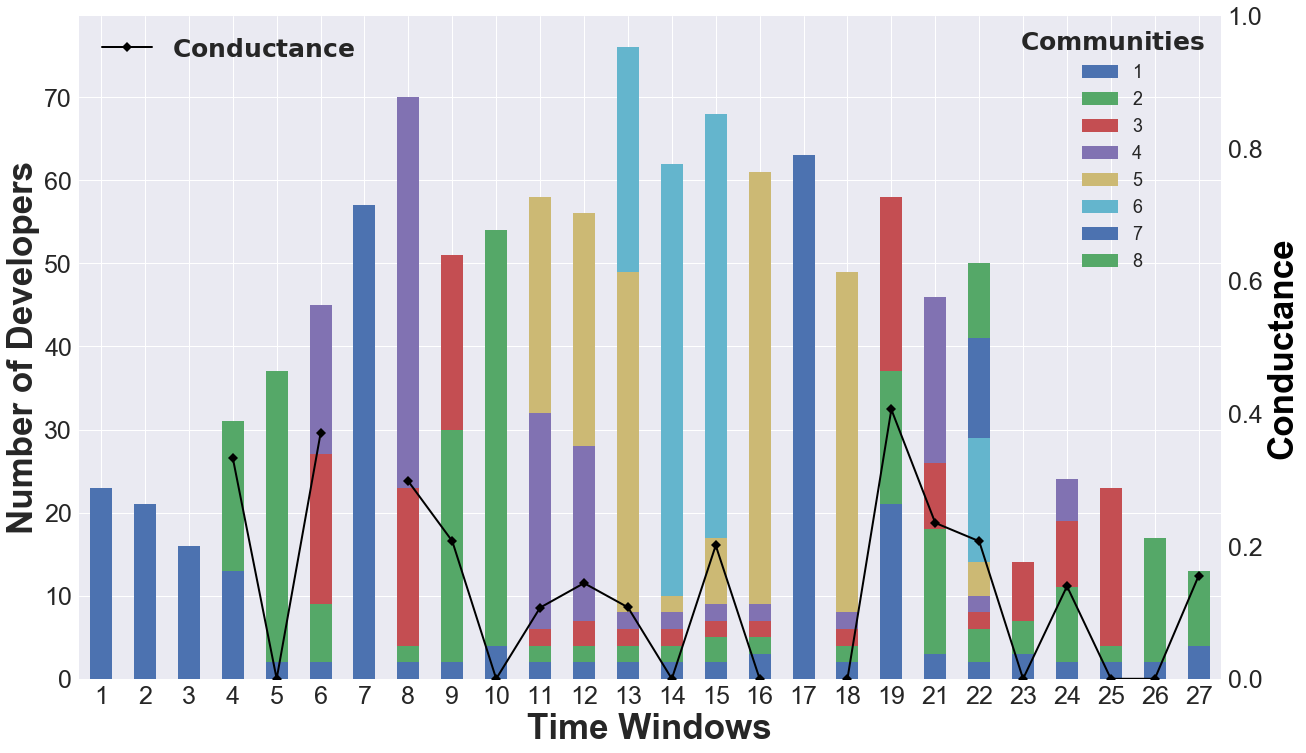}
    \caption{Size of the detected communities and average community conductance for the  \emph{flume} project.}
    \label{fig:example-details}
\end{figure}

For answering [RQ1]: \textit{To what extent can we identify well defined development communities [\ldots]?} we analyse for each project how many communities we find for each time window and their conductance.
Fig. \ref{fig:example-details} displays the count, size, and average conductance of communities found for the example \textit{flume} project across the observed time windows (here twenty six windows (window-20 has less than  10 active  developers therefore not included).
The number of communities ranges between one and eight communities. Note that as communities are latent -- they do not have an explicit identifier but can be observed -- community 4 in window 11 might be most similar to community 5 in window 12 in terms of overlapping members. The size of a community, i.e., number of developers in it, varies as well. As shown in Fig. \ref{fig:example-details}, often communities exceed typical team sizes of ten members (as also previously observed \cite{DBLP:conf/sigsoft/BirdNMGD11,7194606}) as communities also include rarely and non-contributing members. These rarely-contributing members are significant for estimating software quality (e.g., defect prediction) \cite{DBLP:conf/sigsoft/BirdNMGD11}.

\begin{figure}[!ht]
    \includegraphics[width=1.0\columnwidth]{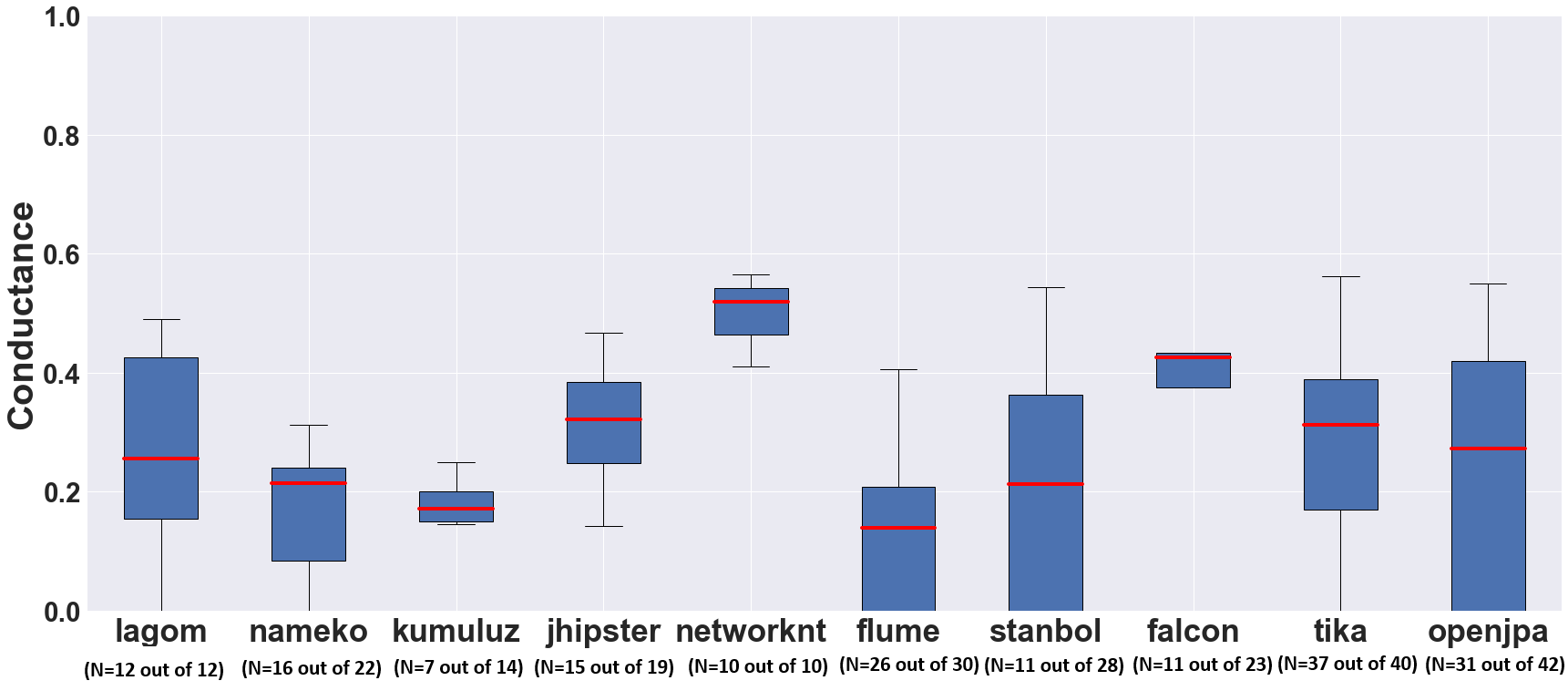}
    \caption{Range of conductance per project. For each project, the x-axis reports also the number of  analysed time windows with at least 10 active developers out of all time windows.}
    \label{fig:projectsconductance}
\end{figure}

\begin{figure}[!ht]
    \includegraphics[width=1.0\columnwidth]{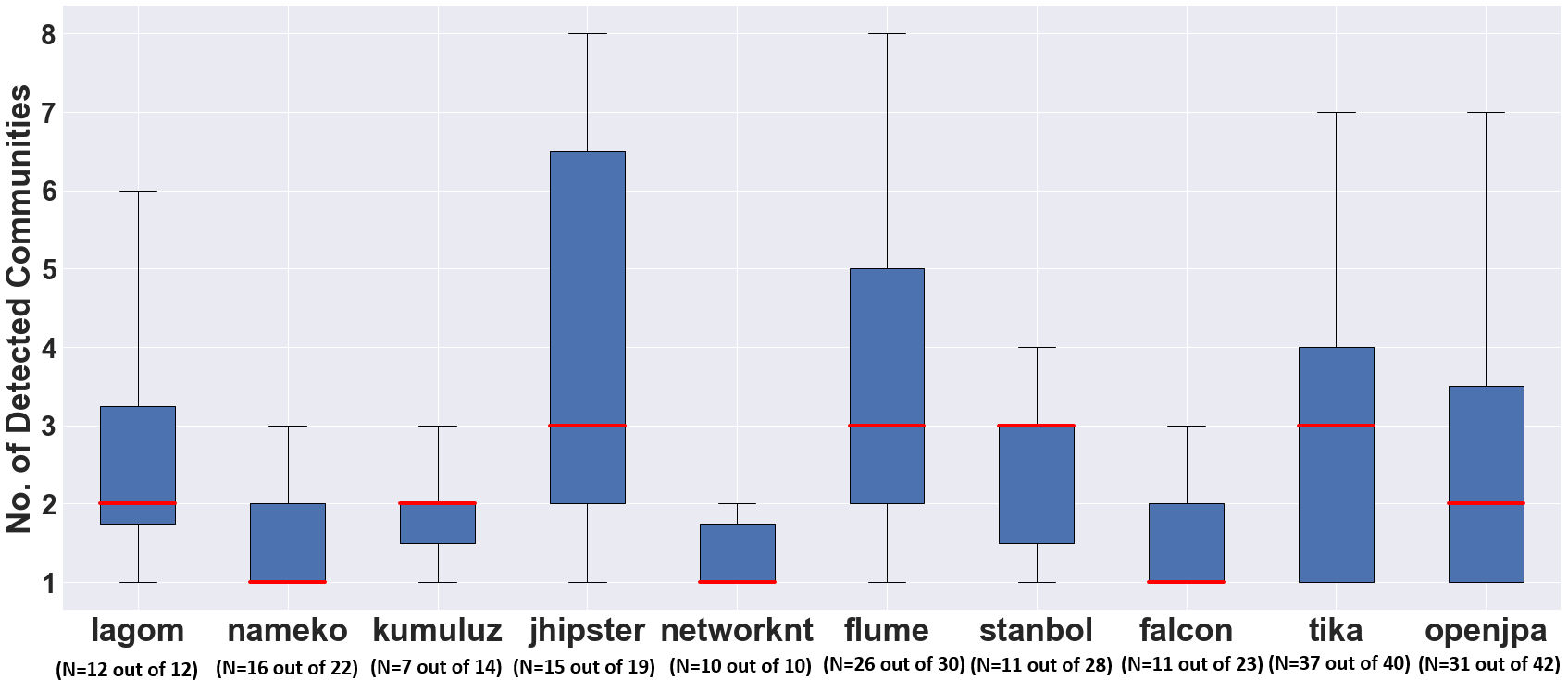}
    \caption{Range of community count per project. }
    \label{fig:communityCountCompare}
\end{figure}

\begin{figure}[ht]
    \includegraphics[width=1.0\columnwidth]{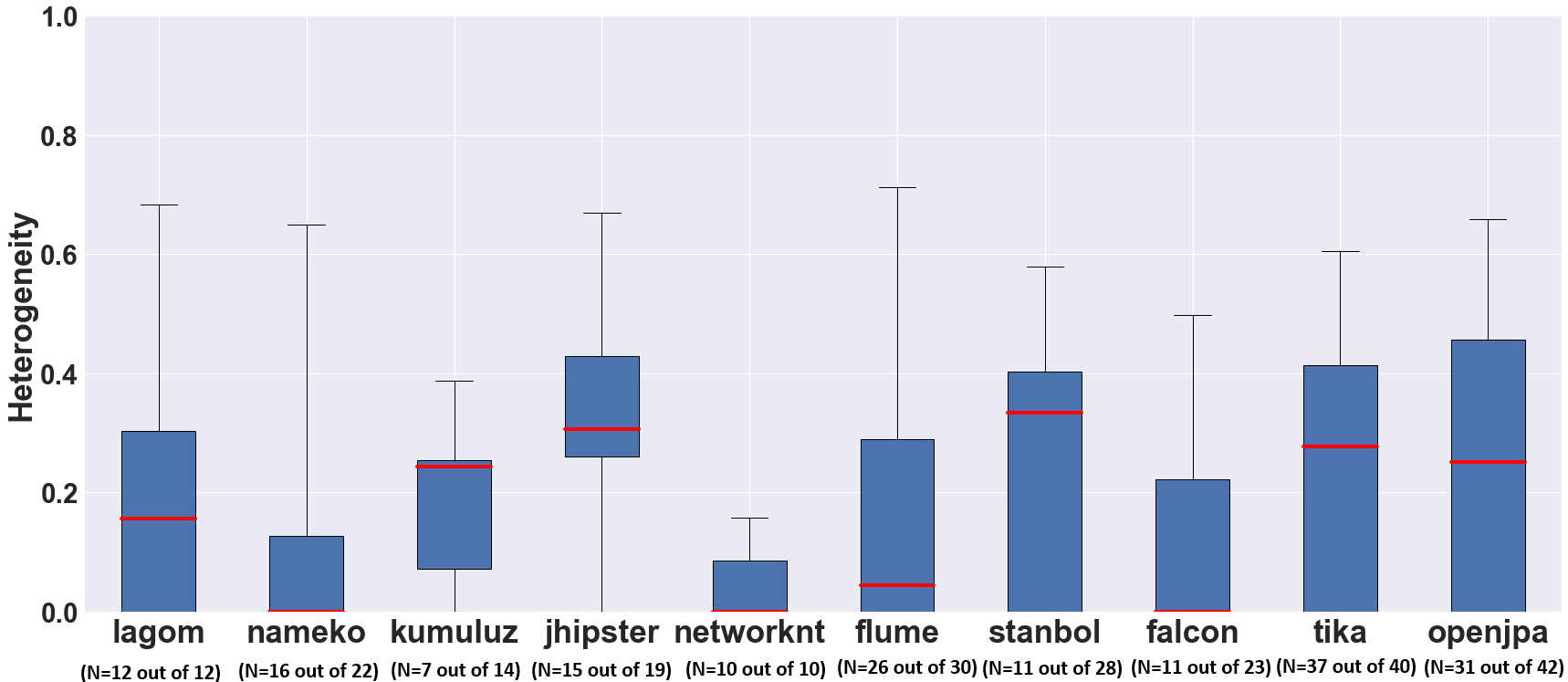}
    \caption{Range of average SDT membership heterogeneity per project.}
    \label{fig:smSDTcompare}
\end{figure}

For comparing the investigated projects, Fig. \ref{fig:projectsconductance} provides a boxplot for each project describing the range of conductance over all time windows that have at least 10 active developers.
We notice that the mean conductance for each project is below 0.4 except \emph{networknt} and \emph{falcon}, with no time window of any project exceeding a conductance of 0.6, a range similar to previous observations for
such a developer count \cite{7194606}.

Fig. \ref{fig:communityCountCompare} reports the range of community count $k$. For all projects the number of community count ranges between one and eight.

\textbf{Observation 1}: There are far fewer communities that emerge from the developer interaction network than subsystems (e.g., \emph{flume} has 16 subsystems).

\textbf{Observation 2}: Overall, communities tend to have medium to low conductance, thus the observed developer interaction networks exhibit a clear community structure.

\begin{figure*}[!ht]
\begin{minipage}{.49\textwidth}

  \includegraphics[width=.99\columnwidth]{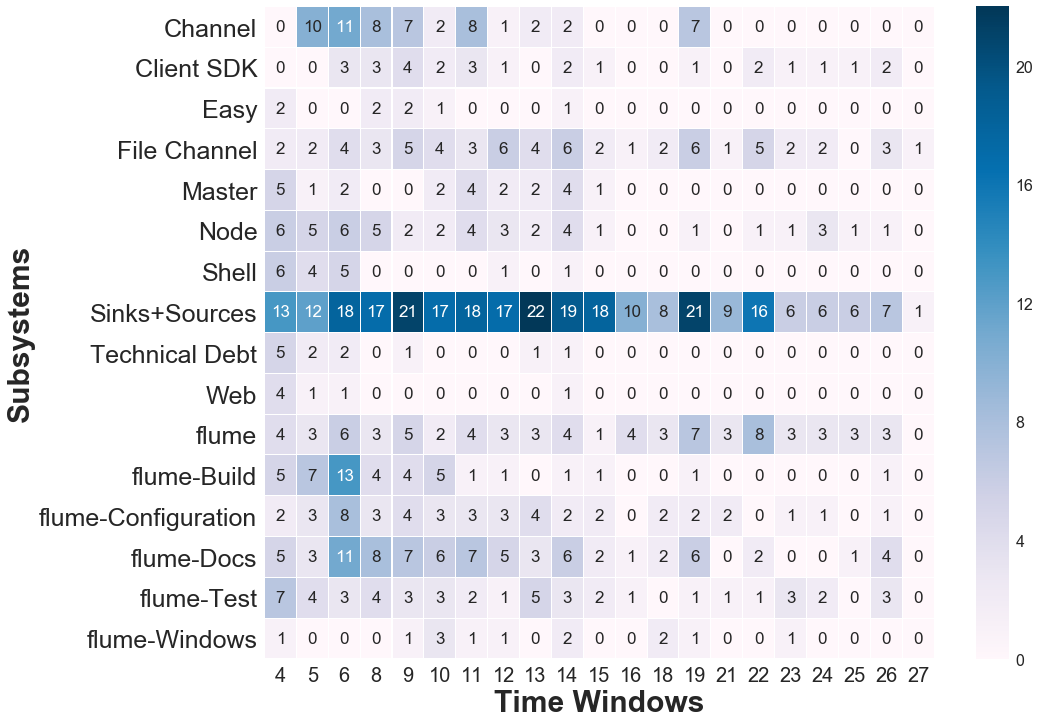}
    \caption{Heatmap of SDT size for the  project \emph{flume} for all time windows with at least 2 communities.}
    \label{fig:SDTexample}
\end{minipage}%
\begin{minipage}{.49\textwidth}

  \includegraphics[width=.99\columnwidth]{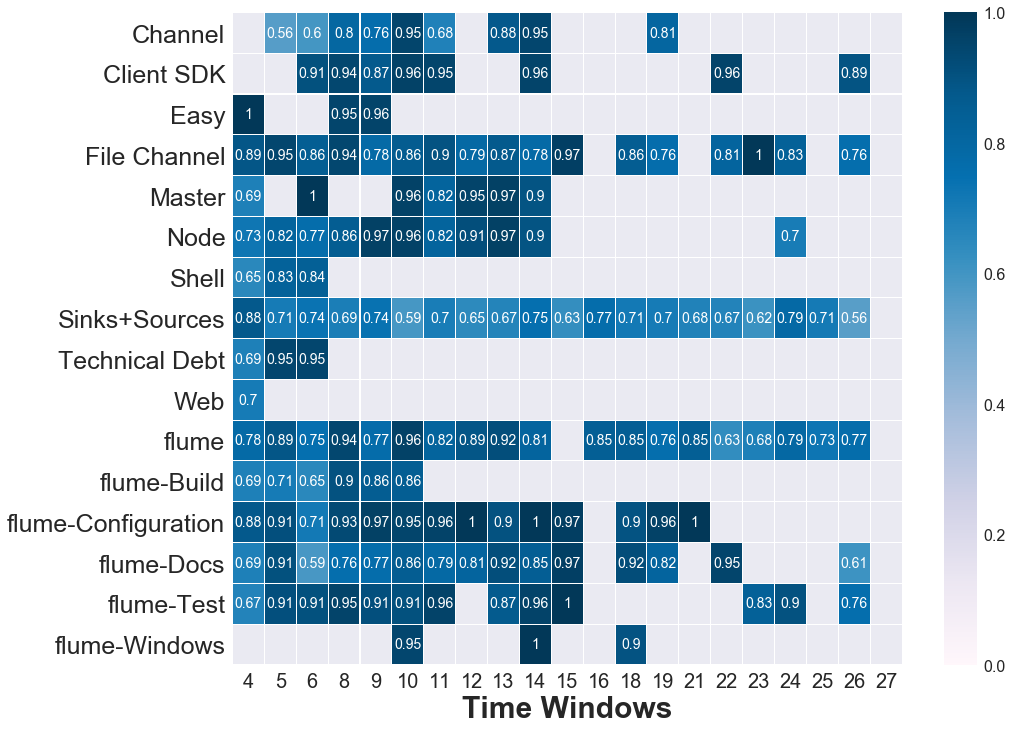}
    \caption{Heatmap of SDT conductance value for each subsystem (with at least 2 active developers) in the project \emph{flume}.}
    \label{fig:smSDTexample}
\end{minipage}
\end{figure*}

\subsection{Answering RQ2}
For answering [RQ2]: \textit{Do the developers active in the same subsystem emerge in the same development communities?} we calculate SDT membership heterogeneity for each subsystem in each observed project for all time windows with at least 10 active project developers and at least two communities (heterogeneity between a single community and SDT is 0).

Fig.~\ref{fig:smSDTcompare} compares the range of average SDT membership heterogeneity across all intervals per project, i.e., we averaged the SDT membership heterogeneity metric over all subsystems where that metric was calculated. 
 The box plots in Fig.~\ref{fig:smSDTcompare} then describe for each project the range of that heterogeneity average when calculated for each time window.

We notice a wide range of heterogeneity behavior across the projects. While \emph{nameko}, \emph{networknt}, \emph{flume}, and \emph{falcon} display low average heterogeneity, \emph{jhipster}, and \emph{stanbol} yield rather high average heterogeneity. The remaining projects yield a medium range of mean average heterogeneity, yet have the occasional time window where average heterogeneity reaches high values, i.e., up to 0.7.

With significant spread of heterogeneity values across subsystems (in \emph{flume}, for example,  between 0.0 and 0.71), we investigate whether high heterogeneity might be correlated with large SDTs (i.e., the more members in a team, the more likely they come from different communities).  
The data show a weak correlation between SDT size and heterogeneity with a Pearson's correlation coefficient value of 0.46. We observe this phenomenon also in the remaining projects.
  
We subsequently investigate whether SDTs form their own, small communities that OSLOM did not detect. 
Such communities may form when developers interact primarily with other developers working on the same subsystem and not so much with developers of other subsystems.
To this end, we treat each SDT as the members of a virtual community and calculate its conductance.  

Fig. \ref{fig:SDTexample} displays SDT size for each subsystem in the example \emph{flume} project. For easier interpretation, conductance values are provided in Fig. \ref{fig:smSDTexample} next to it.
We observe that SDTs exhibit high conductance values (close to 1 and darker in color), thus they are tightly integrated in the developer interaction network and hence do not represent well defined communities. Recall that, a community with no edges amongst its members has conductance of 1 (arguably not really a community at all).

\textbf{Observation 3}: Projects are diverse with respect to how well SDTs typically overlap with a single community. All projects exhibit intervals in which SDTs experience high heterogeneity.

\textbf{Observation 4}: SDTs do not form subcommunities in the developer interaction network but rather maintain considerable interaction ties with developers not involved in the particular subsystem.

\begin{figure}[!ht]
    \includegraphics[width=1.0\columnwidth]{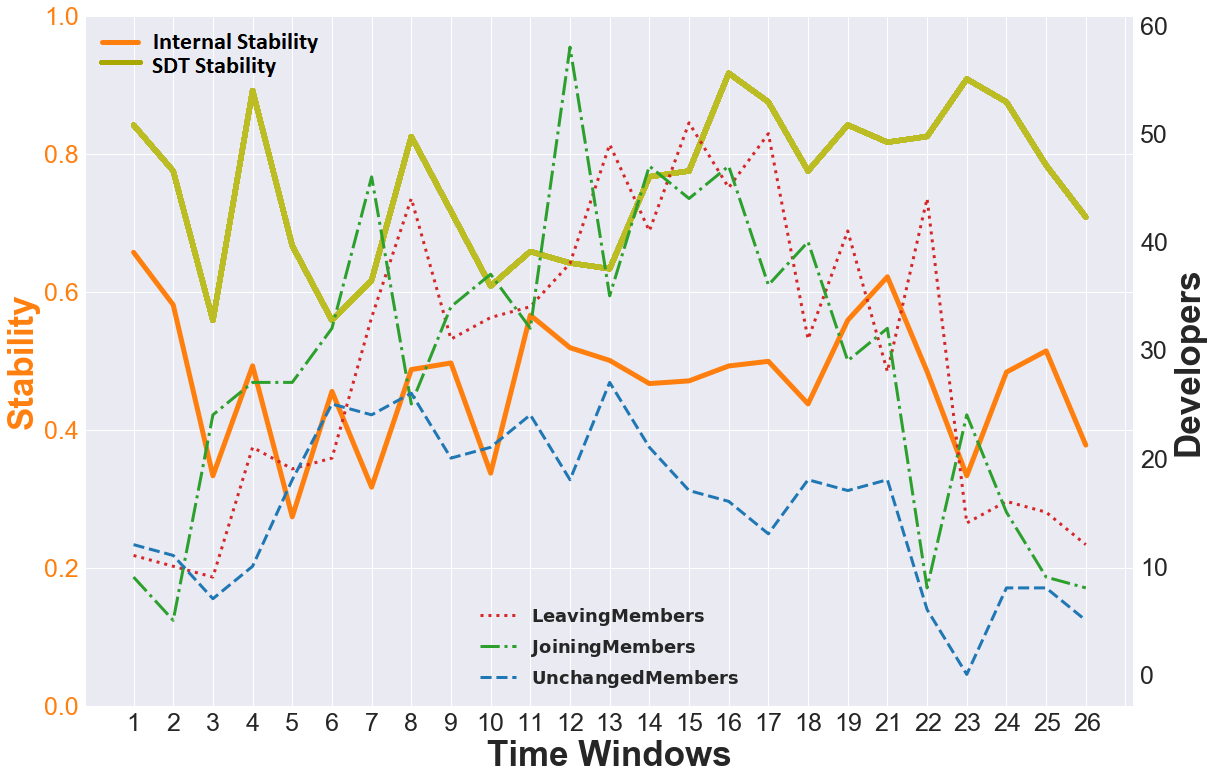}
    \caption{Evolution of Developer Community Membership for example project \emph{flume}: number of developers leaving (dotted line), joining (dash dotted line), and remaining (dashed line) as measured between subsequent intervals (right y-axis), and internal stability as well as SDT stability (left y-axis). }
    \label{fig:iFlucExample}
\end{figure}

\begin{figure*}[!ht]
\begin{minipage}{.5\textwidth}
    \includegraphics[width=1.0\columnwidth]{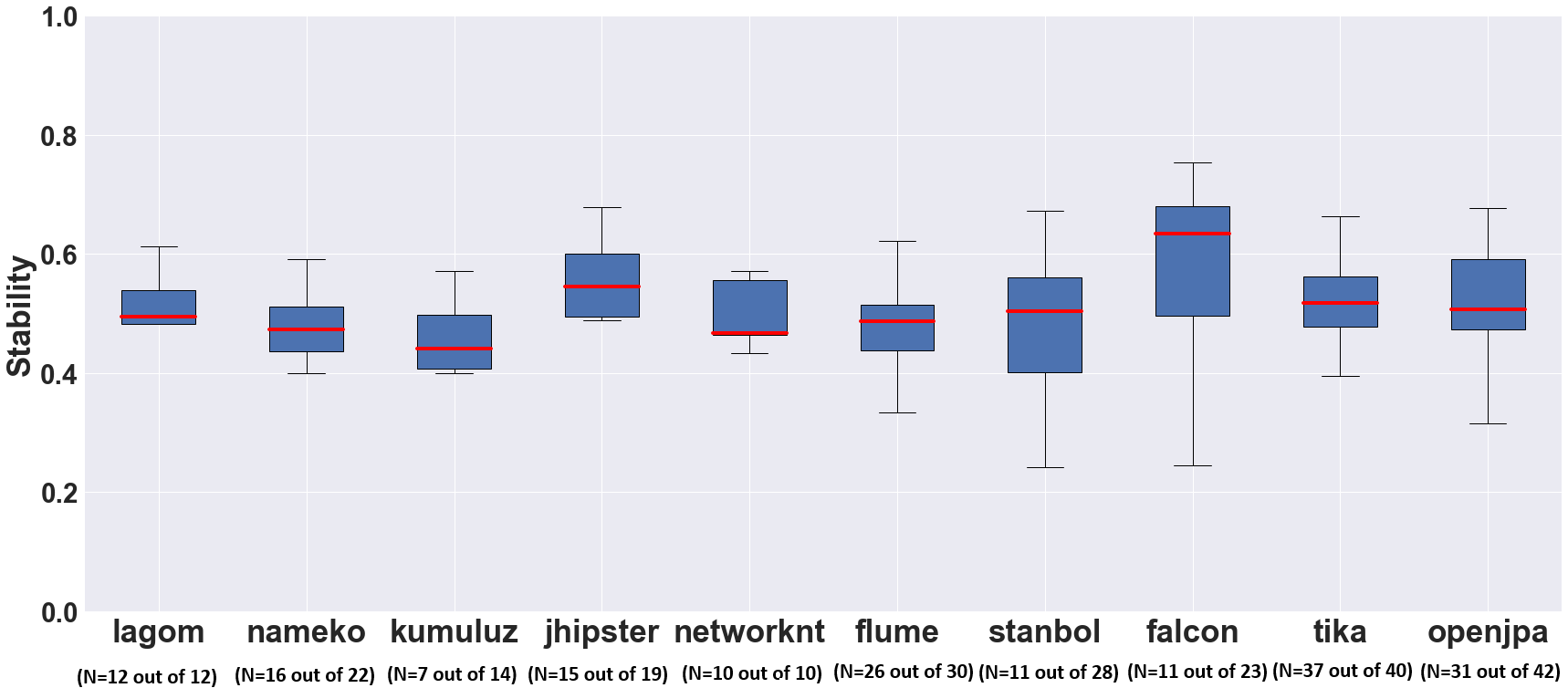}
    \caption{Range of internal stability per project.}
    \label{fig:iFlucCompare}
\end{minipage}%
\begin{minipage}{.5\textwidth}
    \includegraphics[width=1.0\columnwidth]{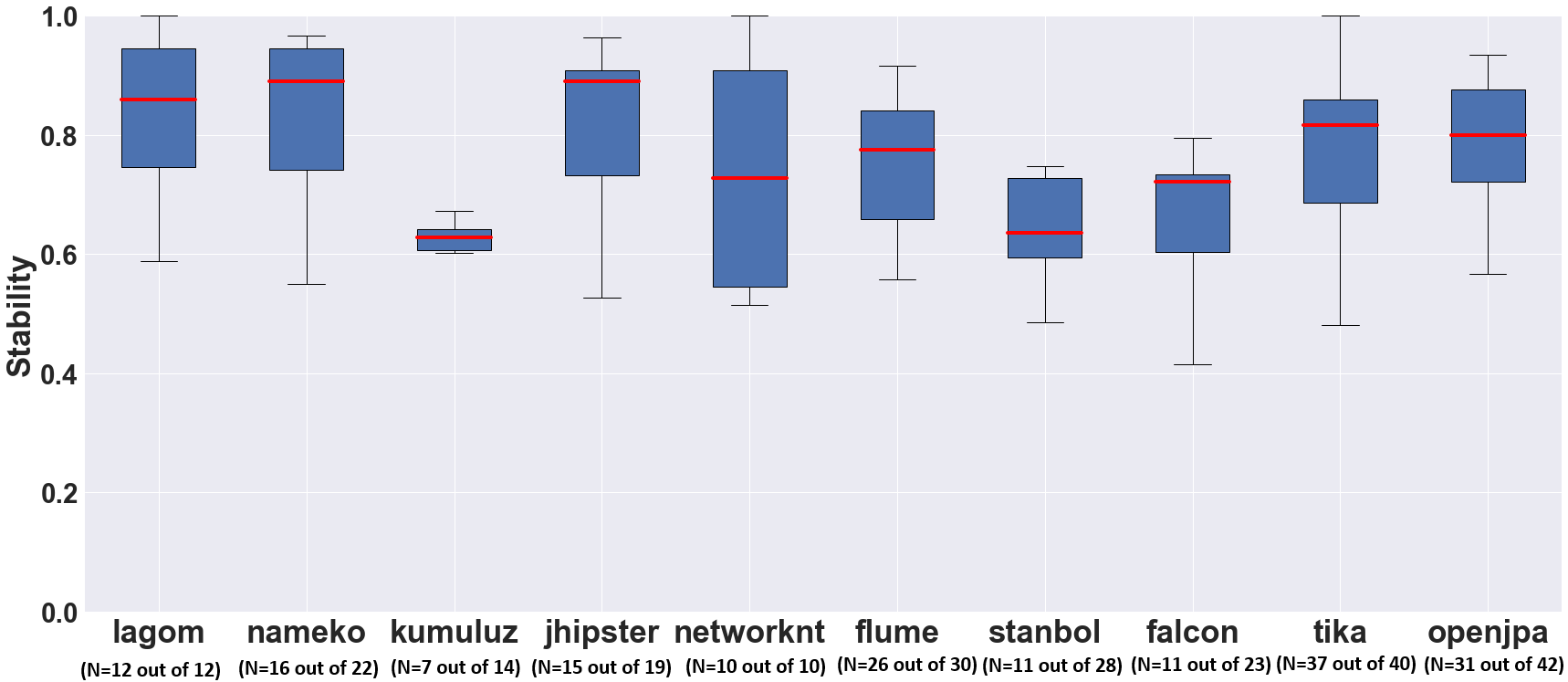}
    \caption{Range of SDT stability per project.}
    \label{fig:SDTFlucCompare}
\end{minipage}
\end{figure*}

\subsection{Answering RQ3}
For answering [RQ3]: \textit{How stable are the detected communities across time compared to the SDTs?} we need to measure internal stability of the communities between two subsequent intervals to understand how much the developer interaction network structure itself changes. We then compare the internal stability with the SDT stability.

Fig. \ref{fig:iFlucExample} reports the evolution of the example project \emph{flume} developer community membership as measured by external and internal stability (interval t reports the change from interval t to t+1).
We notice a heavily fluctuating number of unchanged developers (between 0 to 27)  while an even higher number of  developers keep joining (between 5 and 58) and leaving (between 9 and 51).
This comparatively large set of fluctuating developers influences the community structure, hence few stable key members substantially shift between newly forming communities. This is reflected by the internal stability metric drifting between 0.27 and 0.65. The SDT stability remains above that between 0.65 and 0.91.

For every project we select the time windows with at least 10 active developers. We then determine the internal stability of developers across communities (Fig. \ref{fig:iFlucCompare}). We also determine the stability of SDT overlap with communities (Fig. \ref{fig:SDTFlucCompare}): i.e., whether two SDTs mostly overlapping with the same community in time window $t$ do so also in time window $t+1$.

\textbf{Observation 5}: In general, we find that internal stability is not that strong, hence implying that communities change considerably across time. In contrast, we find considerably stronger SDT stability. This indicates that developers remain in the same SDT but switch among communities (i.e., predominately interact with different members across intervals).

\section{Discussion and Implications} \label{sec:discussion}

The findings of our study are that well defined communities emerge for time windows with $>=10$ developers.
We observe that the alignment between these communities and SDTs covers the complete spectrum from high to low overlap with no discernible influence of the SDT size. Hence, we draw the conclusion that there is no correlation between communities and SDTs: working on a subsystem does not cause interaction links to occur predominantly among SDT members.

We make the careful hypothesis that the low overlap is to mitigate the high developer fluctuation. Interaction links among members from multiple SDTs enable to better cope with situations when developers leave. Then the burden to take over or on-board new members is not placed only with members of that SDT. Cross-SDT interaction may thus be a mechanism to remain robust in the presence of high fluctuation. The low community stability can then be explained as the remaining developers rearrange (at least partially) their interactions and coordination around newly joining members (and the effect of leaving members).

The significant interactions across SDT boundaries also raise the question whether a microservices-centric approach may be suitable in the context of open source software development. The perceived advantages of microservices, among others, are a reduction of coordination needs across microservice-centric teams (i.e., SDTs) \cite{10.1145/3234152.3234191}. Assuming our hypothesis outlined above is correct, the open source community might benefit, even need, such cross-SDT interaction and hence might not be able to benefit fully from a microservice-centric approach. We are not expecting such high overlap in industrial settings -- a subject of future studies.

\textbf{Researchers and Practitioners Implications}
Our work provides a mechanism to measure development communities and their overlap with SDTs. We provide evidence of heterogeneity of development communities and where evolving sub-communities do not overlap with the subsystems compositions. This approach is potentially useful for practitioners, e.g., lead developers in open source projects, to assess the interaction structure among developers and potentially identify subsystems where teams are less well connected. Similarly, practitioners in industrial setting may find the approach useful to identify subsystems where teams are highly interacting, thus perhaps identifying inadvertently strongly coupled subsystems.

\subsection{Threats To Validity}
 \textbf{Construct Validity}
We generated the developer interaction network from issues and commits linked to those issues to identify developers that coordinate their work. We did not contact developers to verify the detected community structure as we do not expect them to accurately recall differences in community structure for fine-granular time windows of 4 months for several years into the past.

Besides issue trackers and commits, we did not consider other possible developer communication channels, i.e., mailing lists, IRC, and conference calls,  which could impact the links created among developers. We investigated that among our analysed projects only Jira-backed projects use mailing lists. Panichella et al. \cite{PanichellaBPCA14} investigated how developers collaboration links vary when data is gathered from different sources. They found, on the one hand, that communication links obtained from mailing lists have high overlap with links obtained from issue trackers and, on the other hand, that emails as the primary communication channel is increasingly replaced by chat and issue trackers. 
Other communication channels such as IRC and conference call are also used in practice. Four of the analysed projects use a chat platform (i.e., \emph{Gitter}\footnote{\url{https://gitter.im/}} where conversations often happen in private threads which cannot be mined). However, research~\cite{PanichellaBPCA14} has shown that mining links from chat is less reliable as 1) this tends to produce too many links, and 2) conversations are less easily associated with issues.   
We therefore believe that we miss only negligible information by leaving out mailing list and chats. We, however, like to point out that co-located developer interactions are not considered and hence restrict the applicability of our approach to projects with completely distributed developers.

\textbf{Internal Validity}
In order to address the internal validity threat, we analyse data from multiple open source systems rather than conducting controlled experiments. The analysis focused on commits and issues in general and
was not specifically tailored to Java projects.
We sampled the chosen projects based on different subsystem structure
mechanisms: mono-repo and multi-repo based to avoid a single mechanism influencing the SDT stability or evolution thereof.

Some of the chosen projects show a low percentage of linkage between commits and issues, thus a threat to the validity of the study. However, we observed similar community quality metric results of 10 additional projects with higher linkage values (the results are a part of the SOM \cite{SOM}). These projects are not included in our study as they do not completely fulfill other inclusion criteria mentioned in the Section \ref{sec:datagathering}. 

We cannot be absolutely certain that the mapping between Jira issue components and Github folders is absolutely correct. A few Github folders for the five Jira-backed projects that could not be mapped to a component with high confidence were excluded to avoid skewing the results.

\textbf{External Validity}
Generalizability of our observations is limited to open source projects that make significant use of issues (incl. pull requests) as their primary coordination and communication mechanism. 
We do not expect our insights to apply to commercial projects where developer teams are top-down defined rather bottom-up emerging.  
The developers in commercial projects are typically (at least partially) collocated. Even when a significant amount of communication happens online, a significant amount of coordination is expected to occur off-line, thus resulting in a less accurate dataset with respect to communication.

\section{Related Work} \label{sec:sota}

Several researchers have analyzed developer interaction data from various sources, such as version control systems, mailing lists, or issue trackers, to investigate how emerging development teams are formed in open source projects.

Joblin et al. \cite{7194606} provide an approach to identify the community structure of a software project. They capture a view on developer coordination, based on commit information and source-code structure. We applied the same algorithm for identifying communities, however, we applied developer interactions based on their involvement in issues along with commit information. Moreover, they capture a macro-level view (once for the overall project) of coordination compared to our approach which focuses on a more fine granular level (developer interaction network alignment with SDTs at separate snapshots in time).

Researchers also examined the effect of interaction between developers on software quality. Tamburri et al. \cite{8651329} \cite{8546762} explored the relation between community smells and code smells in open source environments. They base the detection of community smells on the micro-granular structural differences (i.e., non/existence of edges between individual developers) between the collaboration network generated from commits and the network derived from developer interactions in issues or mailing list. They also propose YOSHI  \cite{DBLP:journals/ese/TamburriPSZ19}, a tool to monitor key community traits in open-source projects. In contrast, our approach observes the community level but not patterns of edges around individual developers.

Bird et al. \cite{DBLP:conf/sigsoft/BirdNMGD11} examined the effect of different ownership measures on release failures in an industrial setting. They found that social network metrics are useful predictors.

Leibzon \cite{DBLP:conf/asunam/Leibzon16} studied the organization of software development teams and project communities at Github. Nzeko’o et al. \cite{DBLP:conf/bigdata/NzekooLT15} made a social network analysis and comparison of developers' and users' mailing lists of four open source software projects. Similarly, Bird et al. \cite{Bird:fse08} analysed the latent social structure by forming a social network from the project's mailing list.
They showed results that sub-communities arise within a project as the project evolves. 

Several efforts aim to exploit the information embedded in the social structure. Canfora et al. \cite{10.1145/1985441.1985463} mined explicitly mentioned cross-system bug resolutions and correlated these activities with social attributes of developers who participated in discussions, e.g., developer mailing lists and commits in source code. Their study showed that cross-system bug fixing mainly involves developers who engage the most in mailing list interaction and developers who are among the top committers. Mockus et al. \cite{mockus2002two} used email archives of source code change history and issue reports to quantify aspects of developer participation, core team size, code ownership, and productivity for open source software projects.

Panichella et al. \cite{PanichellaCPO14} studied how emerging teams evolve over time and that these teams tend to work on more structurally and semantically related set of files. Whether these files belong to or represent a particular subsystem was not part of the study. Hong et al. \cite{ hong2011understanding} also conducted a study to understand developer social network and its evolution. They observed that developers and their relationships change continually. Avelino et al.\cite{10.1007/978-3-319-57735-7_15} studied how authorship-related measures evolve in open-source communities. While they focused only on one huge system, i.e., the Linux kernel, we in contrast focused on several comparatively smaller open source systems.
 
To the best of our knowledge, no approach studied how emerging communities align with subsystems, nor how stable SDTs are over time compared to developer communities.

\section{Conclusion and Future Work} \label{sec:conclusions}
In this paper, we investigate the emergence of latent developer interaction communities and how they align with subsystem developer teams (SDT). We observed that developer community membership is less stable than subsystem developer teams. We noticed hardly any correlation between detected communities and SDTs.  
One cause behind the observed low overlap between communities and SDTs could be the need to remain robust against high developer fluctuation in an open source development environment.

As a future work, we are interested in comparing in more detail those projects that experience different levels of developer fluctuation, respectively what other factors might bring about the lack of correlation between communities and SDTs.

\section*{Acknowledgment}
This work was supported by the Austrian Science Fund (FWF): P31989 and P29415-NBL, and the LIT Secure \& Correct Systems Lab funded by the state of Upper Austria.

\bibstyle{IEEEtran} 
\bibliography{references}

\end{document}